\documentclass[12pt]{iopart}

\usepackage{dcolumn}
\usepackage{graphicx}
\usepackage{booktabs}
\usepackage{color}
\usepackage{bm}
\begin{document}

\title{Modeling mutual feedback between users and recommender systems}

\author{An Zeng$^{1,2}$\footnote{anzeng@bnu.edu.cn}, Chi Ho Yeung$^{3}$, Mat\'u\v{s} Medo$^{2}$, Yi-Cheng Zhang$^{2}$\footnote{yi-cheng.zhang@unifr.ch}}

\address{
$^{1}$ School of Systems Science, Beijing Normal University, Beijing, 100875, P. R. China\\
$^{2}$ Department of Physics, University of Fribourg, 1700 Fribourg, Switzerland\\
$^{3}$ Department of Science and Environmental Studies, The Hong Kong Institute of Education, Hong Kong}

\begin{abstract}
Recommender systems daily influence our decisions on the Internet. While considerable attention has been given to issues such as recommendation accuracy and user privacy, the long-term mutual feedback between a recommender system and the decisions of its users has been neglected so far. We propose here a model of network evolution which allows us to study the complex dynamics induced by this feedback, including the hysteresis effect which is typical for systems with non-linear dynamics. Despite the popular belief that recommendation helps users to discover new things, we find that the long-term use of recommendation can contribute to the rise of extremely popular items and thus ultimately narrow the user choice. These results are supported by measurements of the time evolution of item popularity inequality in real systems. We show that this adverse effect of recommendation can be tamed by sacrificing part of short-term recommendation accuracy.
\end{abstract}
\maketitle

\section{Introduction}
Even if we do not notice it, our life on the Internet is influenced by recommendations. Popular web sites such as Amazon, Netflix and YouTube attempt to facilitate our navigation by suggesting us new possibly relevant items and thus increase our satisfaction and their profits~\cite{e-commerce,WWW200522,JMIS2345}. Employed recommendation algorithms range from simple variants of ``buyers who choose item A also choose item B''~\cite{konstan97,amazon_ICF,adomavicius05} in Amazon to more sophisticated techniques such as the singular value decomposition~\cite{gravity}. Even though many users still act independently of any automated assistance, the use of recommendation is on a rise. For example, the DVD rental company Netflix estimated that 75$\%$ of the rental choices of their users come from some form of recommendation~\cite{netflix2012}.

\begin{figure}
\centering
\includegraphics[scale=0.55]{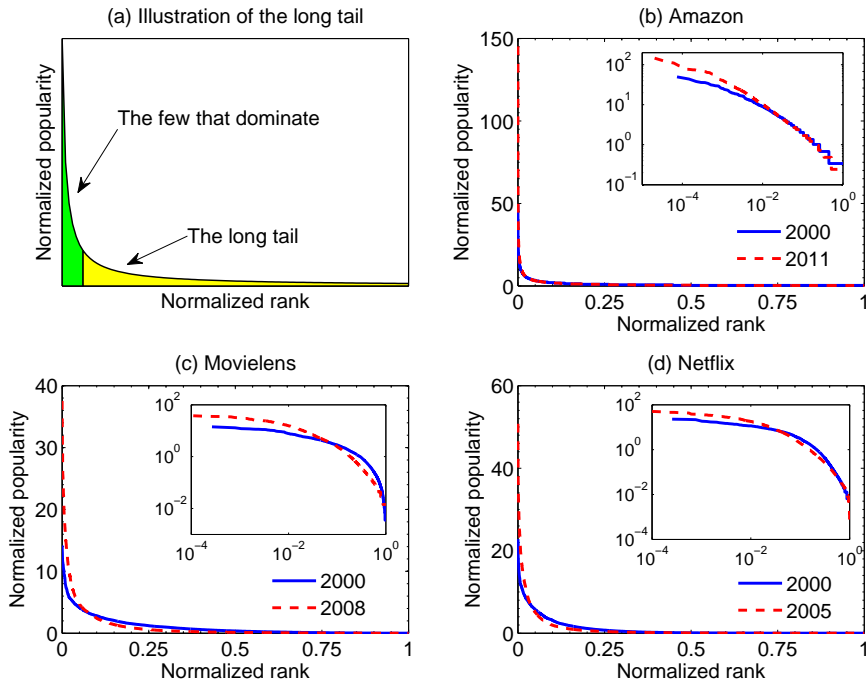}
\caption{(a) A simple illustration of the long tail phenomenon, and (b--d) its presence in real data produced by various e-commerce systems. The normalized item popularity is used to remove the size effect~\cite{normpop}. The same plots in log-log scale are shown in the respective insets. Results from different time periods show that popular items become more popular compared to the niche items at the tail, contrary to the belief of a thicker tail obtained through recommendation algorithms. }
\label{fig:introduction}
\end{figure}

The rationale behind recommendation is to match the right customers with the right products. This task is particularly important and difficult for less popular items for which user patterns cannot be easily identified. Correct matching of little popular items is crucial for e-commerce---studies have shown that from 20$\%$ to 40$\%$ of Amazon's sales do not come from the best selling items~\cite{brynjolfsson06}. It has been suggested that if one ranks items according to their sales and thus constructs a so-called popularity-rank curve (see an example in Fig.~\ref{fig:introduction}a), there is a long tail which comprises a large number of niche items~\cite{anderson06,ACMTW15}. These niche items enjoy a higher profit margin compared to the small profit margin determined by a more competitive market of popular items, and can even boost the sales of other items by providing a convenient one-stop outlet to users~\cite{anderson06,yin12}. In this respect, recommendation algorithms seem to be the best candidate to explore the profit that hides in the long tail.

Recommendation working as intended thus contributes to: (1) Increasing the diversity of recommended items, (2) Distributing user attention more evenly among the items. In this case, recommendation would cause the long tail in Fig.~\ref{fig:introduction}a to gain more weight with time. However, Figs.~\ref{fig:introduction}b-d demonstrate that the opposite is found in reality. Despite recommendation algorithms implemented at Netflix, Amazon, and Movielens (a web site for movie recommendation), the tail of the popularity distribution becomes shorter with time (see also the evolution of the distribution in Fig. S1). Simultaneously, the most popular items account for an increasing share of total sales. It is an adverse effect if some items become too dominant, similar to the emergence of over-dominant species in an ecosystem, which leads to the reduction of biodiversity and ultimately may lead the loss of the equilibrium in the system. Evaluation of recommendation exclusively on the basis of accuracy-oriented metrics cannot, by its nature, capture and explain this long-term behavior. When recommendation is iterated for a small number of rounds as in~\cite{an12}, the possible feedback between user choice and the recommender system is detected. These interactions between users and the system impact on its macroscopic properties and trend, similar to other physical systems. A more physical approach is thus needed to gain the first insights into the puzzle posed by Fig.~\ref{fig:introduction}.

To understand the long-term impact of recommender systems, one has to study the co-evolution of the recommendation and the online user-item network. The recommendation results affect the growth of the network and the change of the network meanwhile influence the future recommendation outcome. The effect is amplified with successive recommendations. In this paper, we investigate this issue by repeatedly applying recommendation on the user-item network. Our focus is different from currently known recommendation studies which aim at short-term metrics such as accuracy~\cite{lu12}, diversity~\cite{PNAS_FR} and others~\cite{herlocker04}. We demonstrate that the repeated use of usual recommendation algorithms makes the system reach a stationary state where user attention is concentrated on a few items instead of distributed over a broad range of items which is in agreement with the data presented in Fig.~\ref{fig:introduction}. In other words, usual recommendation algorithms ultimately narrow user choice and reduce information horizons instead of widening them. We also observe a hysteresis phenomenon which implies that while recommendation naturally gives rise to hugely popular items, to revert this change is not possible because the uneven distribution of user attention is robust over a broad range of the recommendation algorithm's parameters. Our observations directly challenge the role of recommendation for online retailers, as well as other applications of recommendation in search engines~\cite{page99}, online social networks~\cite{lu11}, news media~\cite{kwak10}, and even suggestions of research papers~\cite{sugiyama10}. We finally show that before some items become too dominant, it is possible to make a compromise between recommendation accuracy and long-term effects of recommendation. These results can provide insights and motivations for the design of a next generation of recommender systems.

\section{Methods}

\textbf{Data.}
The data used for the empirical study in Fig. 1 and Fig. S1 is described in the Supplementary Information. In the following simulation, we use two benchmark data sets: Movielens (an online movie rating and recommendation service) and Netflix (a DVD rental service). The Movielens data (available at \emph{www.grouplens.org}) contains $1682$ movies and $943$ users who have rated movies using the integer scale ranging from 1 (worst) to 5 (best). To obtain an unweighted bipartite network, we represent any rating of 3 or more as a link between the respective user and item. The resulting network contains 82520 links. The average degree of users and items is $88$ and $49$, respectively. The maximum degree of users and items is 509 and 558, respectively. The Netflix data is a subset of the original data set released for the purpose of the NetflixPrize (available at \emph{www.netflixprize.com} \cite{netflixprize}). The subset contains 1014 users and 2049 movies chosen from the original data at random and all links among them (since the input data uses the same rating scale as Movielens, the threshold rating of $3$ is again used to determine whether a link is present or not). The resulting network contains 54093 links. The average degree of users and items is $53$ and $26$, respectively. The maximum degree of users and items is 409 and 604, respectively.

\textbf{Models and metrics.}
To model the co-evolution of user choices and the recommendations generated from the recommender systems, we construct a model of recommendation ecosystem as follows. The real data described above are used as the initial configuration. The network evolves through a so-called rewiring process where each link is assigned a time stamp (the initial time stamps are chosen at random). In each rewiring step, the oldest link of every user is redirected to a new item and assigned with the current time (i.e., it becomes the user's newest link). Deleting the oldest links simulates the case where the recommendation results are generated based on recent historical record~\cite{CACM2010}. We assume that when a new target item for a currently rewired link is being chosen, the user follows recommendation with probability $p$ and decides independently of recommendation with the complementary probability $1-p$. We use the widespread Item-based Collaborative Filtering (ICF) as the recommendation algorithm (see details below); variants of this algorithm are employed by Amazon and other major web sites~\cite{amazon_ICF}. When a user follows recommendation, they select an item from their current recommendation list with probability inversely proportional to item rank in the list (the motivation to use rank-reciprocal rather than equal probability for all listed items comes from~\cite{Lempel03,Joachims2007}). When a user acts independently of recommendation, they either choose an item at random (which we refer to as Random Attachment, RA) or they choose proportionally to item degree increased by one (which we refer to as Preferential Attachment, PA; item degree is incremented by one to make it possible for items which lose all their links to be chosen again). As the network evolves, degree of each user is preserved. On the other hand, network structure and item degree values change significantly by rewiring. Network structure can be at any moment represented by a so-called network adjacency matrix $\mathsf{A}$ whose element $a_{i\alpha}$ is one when user $i$ is currently connected with item $\alpha$ and zero otherwise.

We use the Gini coefficient $G$ to measure inequality of the item popularity distribution during the network evolution. While this quantity has been originally proposed to quantify inequality of income or wealth distribution~\cite{Gini_index}, it has also been used in other fields~\cite{Nature458623,PHM37,Min4855}. The Gini coefficient can be computed as
\begin{equation}
G = \frac{2\sum_{\alpha=1}^M\alpha k_\alpha}{M\sum_{\alpha=1}^Mk_\alpha}-\frac{M+1}{M},
\end{equation}
where $k_{\alpha}$, the popularity of objects, has been sorted in the ascending order. The two extreme values of the Gini coefficient are $0$ and $1$ which correspond to equal popularity of all items and zero popularity of all items but one, respectively. An increase of the Gini coefficient thus corresponds to the item popularity distribution becoming more unequal. We also used other measures, such as the link share of 1$\%$ most popular items and the Herfindahl index, to study the rewiring model. Results obtained with different inequality metrics are consistent with those obtained with the Gini coefficient.

\textbf{Recommendation.}
We use the widely spread Item-based Collaborative Filtering (ICF) as the main recommendation algorithm~\cite{ACM_CF}. In ICF, the recommendation score of an item is computed based on the item's similarity with other items collected by a target user. The score of item $\alpha$ for user $i$ reads
\begin{equation}
f^{(i)}_{\alpha}=\sum_{\beta=1}^{M}s_{\alpha\beta}a_{i\beta}
\end{equation}
where $s_{\alpha\beta}$ is the similarity of items $\alpha$ and $\beta$, and $a_{i\beta}$ are elements of the network's adjacency matrix. We choose the following formula for item similarity
\begin{equation}
\label{simil}
s_{\alpha\beta}=\frac{\Gamma_{\alpha}\cap\Gamma_{\beta}}{(k_{\alpha}k_{\beta})^{\theta}}
\end{equation}
where $\Gamma_{\alpha}$ denotes the set of users who have collected item $\alpha$ and $k_{\alpha}$ denotes the degree of item $\alpha$. Parameter $\theta$ can be continuously adjusted in the range $[0, 1]$ which includes three classical cases: the common neighbor similarity which simply counts the number of users who have collected both items (when $\theta=0$), the cosine similarity which is sometimes referred to as the Salton index (when $\theta=1/2$) and the Leicht-Holme-Newman similarity (when $\theta=1$)~\cite{lp_survey}. Due to a lack of normalization, high-degree items are favored when $\theta=0$. By contrast, low-degree items are favored when $\theta>1/2$. Equation~(\ref{simil}) thus gives us the opportunity to gradually move from recommendations biased towards high-degree items (when $\theta=0$) to diversity-favoring recommendations biased towards low-degree items (when $\theta=1$). To obtain a recommendation list for a given user, all items that are currently not connected with this user are sorted according to their recommendation score in a descending order and finally the top $L$ items are kept on the recommendation list. We use $L=20$ here which is a common value in previous studies of recommender systems~\cite{lu12}. Considering only the top $L$ items is motivated by the fact that users in real online systems do not have time to inspect the list of all items ranked by their recommendation score (as further reflected in the rank-reciprocal probability in our link rewiring process).

\section{Results}
\begin{figure}
\centering
\includegraphics[scale=0.55]{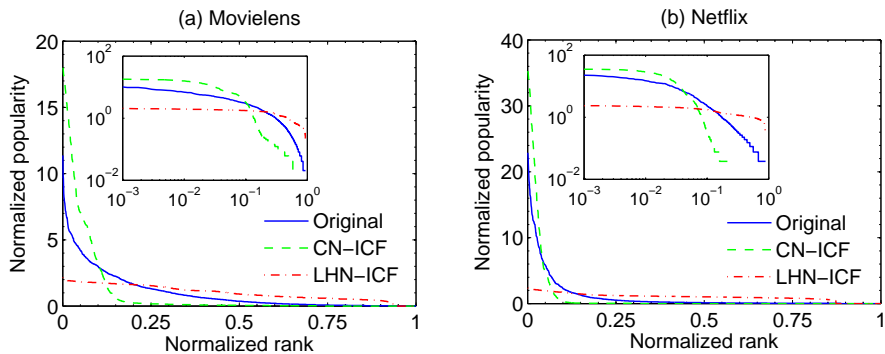}
\caption{The normalized item popularity versus the normalized item rank in (a) Movielens and (b) Netflix. The same plots in log-log scale are shown in the respective insets.}
\label{fig:deg_distr}
\end{figure}

\textbf{Distribution of item popularity.} After a sufficient number of iteration steps in our model, the system reaches a stable state and the item degree distribution becomes stationary. We begin our analysis by comparing the distribution of item degree in the original data with the stationary outcome of the rewiring procedure. We assume here for simplicity that the users choose new items solely by recommendation (i.e. $p=1$). Our implementation of item-based collaborative filtering employs a user similarity metric with one parameter given by Eq.~(\ref{simil}). Three particular cases of our user similarity lead to well-known similarity measures: $\theta=0, 1/2, 1$ correspond to Common Neighbor similarity (CN), Cosine similarity (COS), and Leicht-Holme-Newman similarity (LHN), respectively. The present parameterization allows us to continuously tune between recommendation that favors high-degree (when $\theta$ is small) and low-degree (when $\theta$ is big) items. This is well demonstrated by Fig.~\ref{fig:deg_distr} which shows the item degrees in the original data compared to those after the rewiring procedures using CN and LHN. ICF with CN (CN-ICF) improves the popularity of the most popular items and essentially eliminates the long tail. This inferior outcome produced by an otherwise well accepted and popular recommendation method demonstrates the potential danger of recommendation for information diversity, similar to the loss of biodiversity in an ecosystem. On the other hand, LHN-ICF strengthens the long tail but, as we shall see later, its recommendation accuracy is low.

To obtain a quantitative comparison, we measure the Gini coefficient over the item degrees. The higher the values, the more uneven the distributions, and the greater the loss of information diversity in the system. The Gini coefficient corresponding to the distributions shown in Fig.~\ref{fig:deg_distr} are $0.31$, $0.67$, $0.88$ (Movielens) and $0.28$, $0.82$, $0.95$ (Netflix) for the LHN-ICF, original data, and CN-ICF, respectively. The LHN-ICF method leads to a remarkable equalization of the item popularity, while the CN-ICF method further advances the degree heterogeneity. Here we use $L=20$. We show in Fig. S2 for the results when other $L$ values are used. Moreover, we study in Fig. S3 the case where users are influenced by the similarity constraint when selecting items from the recommendation list. We want in this way to prevent all information from being washed away.

\begin{figure}
\centering
\includegraphics[scale=0.55]{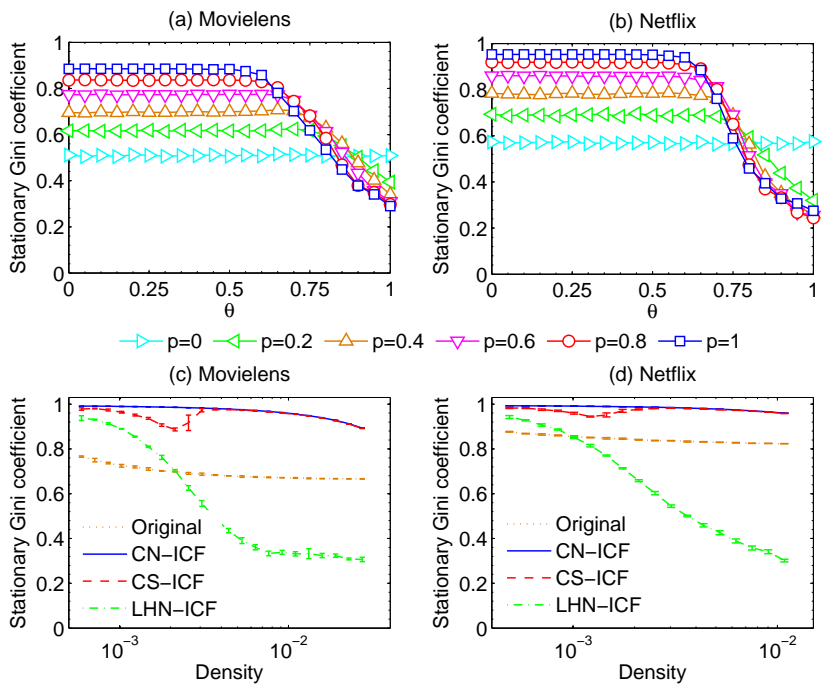}
\caption{(a,b) The effect of the ICF's parameter $\theta$ on the stationary Gini coefficient; links which are not drawn based on recommendation are drawn based on preferential attachment. (c,d) The effect of data density on the stationary Gini coefficient for different recommendation methods (here all links are drawn based on recommendation). The curve labeled original is the Gini coefficient of the network after density modification and before rewiring.}
\label{fig:stationaryG}
\end{figure}

\textbf{User reliance on recommendation.} In practice, users do not always follow recommendations. We thus consider the case with $p<1$, i.e. users follow recommendation with a probability $p$ and otherwise (with a probability $1-p$) choose an item according to preferential attachment (as shown in Fig. S4 and S5, the resulting behavior is similar if preferential attachment is replaced by random or real choice of items). The value of $p$ thus characterizes users' reliance on recommendation, and parameter $\theta$ controls the popularity bias of recommendations. While users' recommendation lists are populated mostly with popular items when $\theta$ is large, the number of niche (little popular) items increases when $\theta$ is small. We repeatedly rewire the input data and again quantify the diversity of the system by the stationary value of the Gini coefficient, $G^*$. As shown in Fig.~\ref{fig:stationaryG}a-b for various $p$ values, $G^*$ is not sensitive to $\theta$ over a broad range $[0,\tilde\theta]$ where $\tilde\theta\approx0.6$. Once $\theta>\tilde\theta$, $G^*$ decreases quickly and eventually reaches values lower than those produced by preferential attachment only. Another important finding out of our expectations is that, recommendation can hurt information diversity even more than preferential attachment (i.e., ICF can lead to higher $G^*$ than PA). This is because PA is not personalized and biased on popularity, but any items may be chosen; recommendation algorithms are personalized, biased on popularity and only the top $L$ items identified by the algorithms are recommended. If the top $L$ items for different users significantly overlap, they will attract a large amount of links. This strong tendency of recommendation to decrease information diversity is a joint outcome of a popularity-favoring recommendation method and the fact that the users are recommended with a short list of items which further contributes to the ``winner takes it all" situation. We also consider a different recommendation algorithm~\cite{PNAS_FR} which show remarkable similarity with the results for ICF presented in this paper (See results in Fig. S6).

\textbf{The impact of data density.} The high density of these two data sets enables us to adjust the density of the network by removing some links. In Fig.~\ref{fig:stationaryG}c-d, we study the effect of data density on the resulting popularity inequality when $p=1$ (see Fig. S7 for the results with $p<1$). we randomly remove links from the original network until the desired density is reached. Note that the last link of the users will never be removed in this process. The effect of data density is particularly strong when recommendations are made with the diversity-favoring LHN-ICF method which leads to low $G^*$ when the data density is high but does the opposite when the data density is low. This can be explained by the presence of objects with only a few links in low-density data: once those links are rewired to other objects, the respective object cannot be any more recommended and it effectively disappears from the system, thus contributing to an increased popularity inequality and a high $G^*$.

Nevertheless, a decrease in the data density does not always increase Gini value. These observations may come from an interesting phenomenon, which can be shown by two different simulations. In the first case, users are randomly removed from the system and $G^*$ is found to increase with decreasing data density (see Fig. S8(c)). In the second case, items are removed at random, and by contrast $G^*$ decreases with decreasing data density (see Fig. S8(d)). While these two scenarios look similar, the average item degree decreases in first case (see Fig. S8(a)) and is preserved in the second case (see Fig. S8(b)). In other words, even though data density decreases, recommendation algorithms work equally effectively to distribute the popularity given the average item degree is preserved. This may come from the fact that the nature of an item can be reflected by the characteristics of individual users who have collected it, the degree of an item represents the amount of available information on it. As a result, when the average item degree is preserved, recommendation algorithm works effectively since there are sufficient information on the items. These results show that the degree of an item is not just merely related to its role in the network, but may also represents the amount of information we possess on it. The effectiveness of recommendation algorithms are thus strongly dependent on the average item degree.

\begin{figure}
\centering
\includegraphics[scale=0.55]{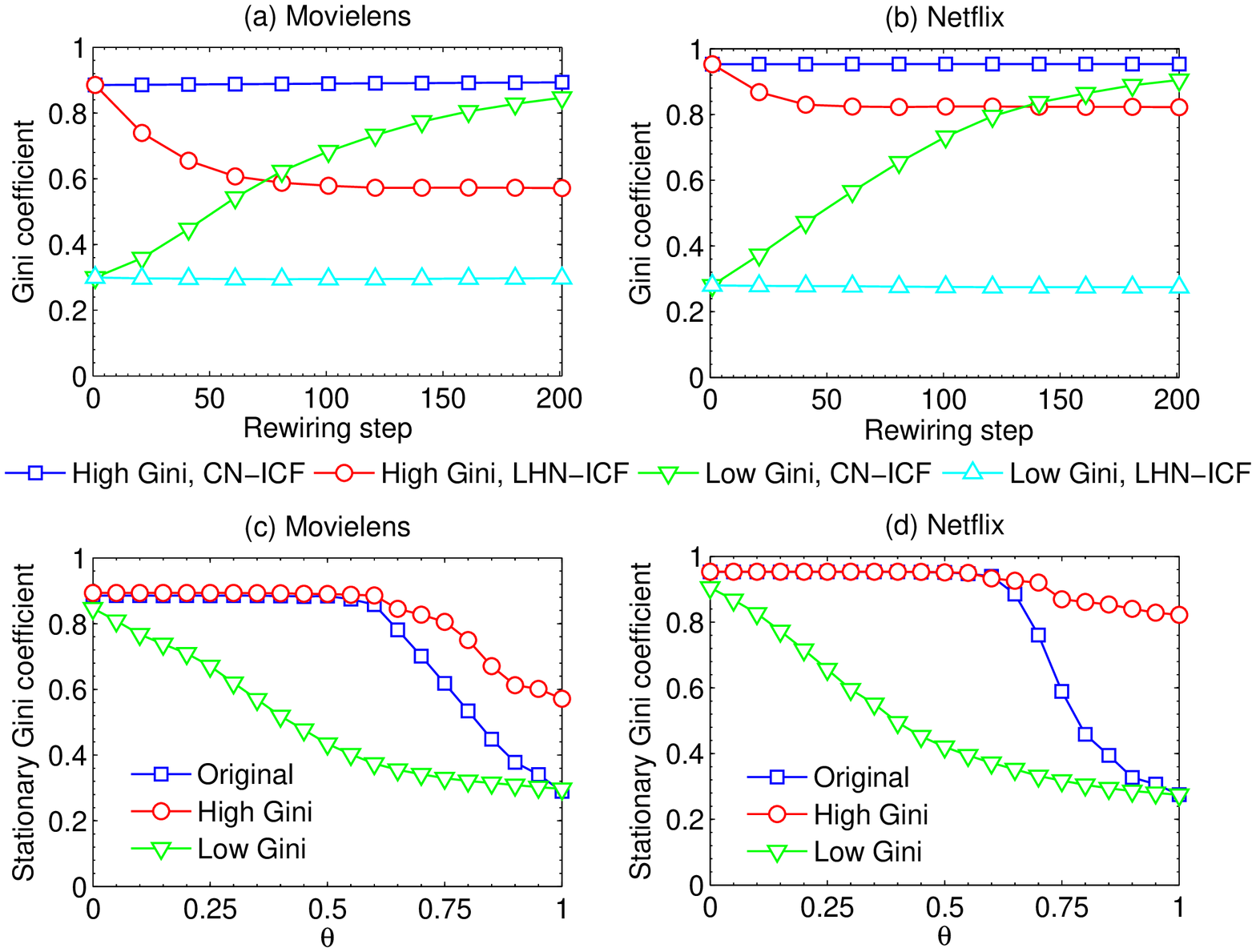}
\caption{Evolution of the Gini coefficient starting from different initial configurations, low-$G^*$ achieved with $\theta=1$ and high-$G^*$ achieved with $\theta=0$, in (a) Movielens and (b) Netflix data. (c) and (d) show the effect of $\theta$ on the stationary Gini (ICF method) under different initial configurations.}
\label{fig:hysteresis}
\end{figure}

\textbf{Hysteresis.} The task of choosing the recommendation parameter $\theta$ is made more important by a pronounced hysteresis phenomenon. Fig.~\ref{fig:hysteresis} shows that while a state with low inequality achieved with $\theta=1$ can be fully reverted to a high inequality state by changing $\theta$ to $0$, the opposite is not true. In other words, once high concentration of item popularity has set in, it can only be partially corrected by the use of a diversity-favoring recommendation method. This is also related to the amount of information available on the objects -- after we lose information on some objects in a recommendation algorithm, we may not retrieve it again. Effective recommendation cannot be made for these items and the popularity remains unevenly distributed, even with diversity-favoring methods. In Fig. S7, we show that the hysteresis phenomenon exists also when $p<1$. We thus conclude that $G^*$ depends on the system's initial condition, especially when it corresponds to highly heterogeneous item popularity.

\textbf{Trade-off between diversity and accuracy.} The previous figures show that the use of high $\theta$ can limit or even reverse the potential popularity-concentrating effect of recommendation. However, high $\theta$ generally leads to low recommendation accuracy because it tends to recommend objects of low degree~\cite{PNAS_FR}. This motivates us to investigate the trade-off between the possibly decreased stationary Gini coefficient $G^*$ and recommendation precision $P$ measured on the original input data before rewiring. To measure recommendation accuracy, we apply the standard evaluation procedure which is based on randomly dividing the network data into two parts: a training set $E^T$ comprising 90\% of all links and a probe set $E^P$ comprising the rest. The training data $E^T$ is then used to compute recommendation lists for all users. The standard accuracy metric called \emph{precision} is based on comparing these recommendation lists with the probe data $E^P$: a good recommendation algorithm is expected to be able to reproduce a large part of $E^P$ based on $E^T$~\cite{herlocker04}. If for user $i$, there are $d_i(L)$ probe entries related to this user in $i$'s recommendation list, we say that the recommendation precision for this user is $P_i(L):= d_i(L) / L$. By averaging this quantity over all users with at least one entry in the probe set $E^P$, we obtain the overall recommendation precision $P(L)$. To further remove its possible dependence on the data division, we average precision over ten independent training set-probe set divisions. Based on the training-probe sets division, one can also measure the short-term recommendation diversity which is simply the average degree of items that appear in the recommendation lists. Fig. S9 shows the relation between the long-term Gini coefficient and short-term recommendation diversity.

\begin{figure}
\centering
\includegraphics[scale=0.55]{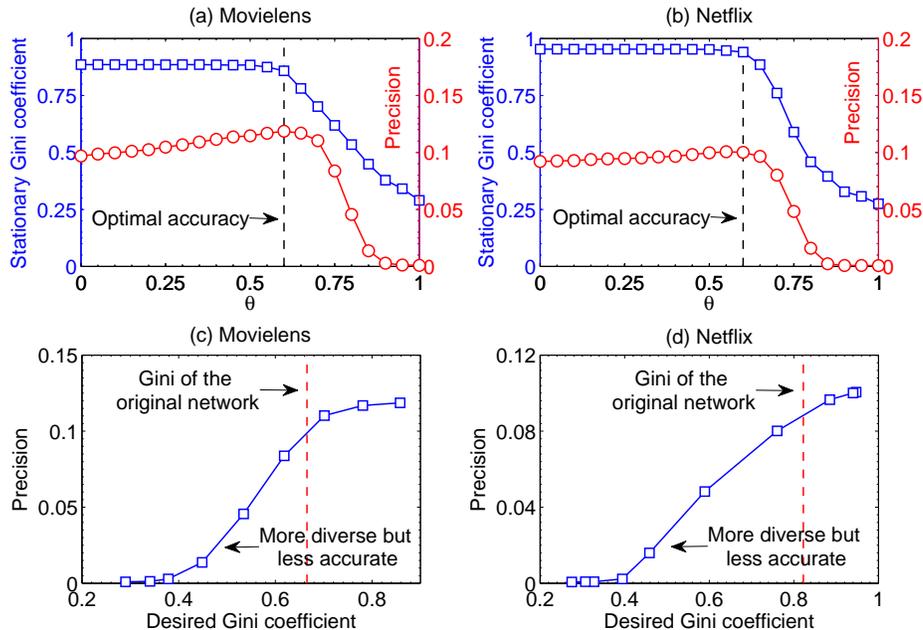}
\caption{The dependence of the stationary Gini values and recommendation precision on $\theta$ in (a) Movielens and (b) Netflix data. Panel (c) and (d) show the relation between the desired stationary Gini coefficient and the short-term recommendation precision.}
\label{fig:sp_hy}
\end{figure}

Fig.~\ref{fig:sp_hy}a simultaneously plots $G^*$ and $P$ as a function of $\theta$ in the Movielens data. One can see that the highest precision is achieved at $\theta\approx0.6$ which is a point where the stationary Gini decreases quickly with $\theta$. This gives us the possibility to lose some precision by increasing $\theta$ from its optimal value and in turn achieve a substantial decrease of $G^*$. This is demonstrated by Fig.~\ref{fig:sp_hy}c where the desired value of $G^*$ is plotted on the horizontal axis: the resulting precision first decreases rather slowly from its highest value as the desired stationary Gini coefficient is lowered. Only when the desired $G^*$ is lower than the Gini value in the original data, precision decreases sharply. The situation is less favorable for the Netflix data where precision is maximized at $\theta\approx0.57$ which lies in the region where $G^*$ changes rather slowly with $\theta$ as shown in Fig. 5b. As a result, one has to substantially increase $\theta$ in order to achieve a significant decrease of $G^*$. This manifests itself in Fig.~\ref{fig:sp_hy}d which lacks the gentle slope region seen in Fig.~\ref{fig:sp_hy}c. Nevertheless, one can limit $G^*$ to the values seen in the original Movielens and Netflix data by sacrificing 17\% and 12\% of the optimal recommendation precision, respectively. These results show the possibility to compromise recommendation accuracy and long-term impacts on diversity for recommendation systems.

\section{Discussion}
By studying the co-evolution of the recommendations generated from the recommender systems and users' choices, we demonstrate the long-term effect of recommendation on the distribution of item popularity. This novel approach to the evaluation of recommendation performance gives us the possibility to observe new phenomena. Contrary to the common belief that recommendation helps to match niche items with users who may appreciate them and thus contribute to improving their recognition, we show that typical recommendation methods reinforce the position of already popular items at the cost of niche items. This is particularly true when recommendation algorithms optimized for their accuracy are used because these tend to favor popular items. Our observations suggest that recommendation may divert the system to a state where a few items enjoy extraordinarily high levels of user attention. Furthermore, we found a strong hysteresis effect which implies that this state is very robust and can be hardly changed back to a state with a more even popularity distribution, even with the help of a diversity-favoring recommendation algorithm of lesser accuracy. All these are adverse effects if one considers the interaction between users and the recommender system as an information ecosystem. We remark that although it is not ideal for buyers to receive recommendations of common popular items, such systems may still benefit the sellers by lowering the effective number of distinct items that they need to keep in their inventory and thus reducing the costs for logistics.

To the best of our knowledge, our model is the first one to analyze the long-term influence of recommendation on the evolution of online systems. We tested many different choices of our model including examining different original data sets and another recommendation method, combining recommendation with random attachment instead of preferential attachment, preventing the real information from being washed out by repeated rewiring, and setting different lengths of the recommendation list. Our finding of the adverse effect of a sub-optimal recommendation system on information ecosystems in the long run is in general still valid in these cases and thus needs to be seriously considered in practice. Our work raises a number of questions which aim to further strengthen our understanding of the long term influence. For example, the model now assumes that all users accept the recommendation with the same probability. One can actually instead consider a scenario where experienced users search for items on their own and thus depend less on recommendation. We measure recommendation accuracy before the rewiring process. It would be interesting to monitor the recommendation accuracy during the network's evolution, and couple it with the rate at which users accept it (the higher the accuracy, the higher the probability that users follow recommendation). These more complicated scenarios may make the results of the model quantitatively different from the current ones. A systematic study on these variances would be an interesting and important extension of the current model.

Our findings suggest a need for a next generation of recommender systems which would take into account both short-term and long-term goals. One might argue that the prime goal of a commercial system is to increase the profit by maximizing the recommendation accuracy and the long-term goal of enhancing or at least preserving the item diversity is secondary. Our results suggest that this is a short-sighted approach as focusing on short-term performance of recommendation may ultimately lead to a system where the long tail has been decimated together with its economic potential. To overcome this problem, item diversity can be enhanced by sacrificing a small fraction of recommendation's short-term accuracy in exchange for higher long-term diversity. A detailed investigation of various approaches to study the long-term effects of recommendation as well as possible trade-offs between short- and long-term performance of recommendation are of great interest to both researchers and practitioners in the future.

\section*{Acknowledgments}
This work was partially supported by the Youth Scholars Program of Beijing Normal University (grant no. 2014NT38), EU FP7 Grant 611272 (project GROWTHCOM) and by the Swiss National Science Foundation (grant no. 200020-143272). The work of CHY is partially supported by the Internal Research Grant RG 71/2013-2014R of HKIEd.


\begin{thebibliography}{99}
\bibitem{normpop} We first denote by $R_\alpha(y)$ the rank of item $\alpha$ according to their increase in degree within a year $y$. We further denote the number of active items in year $y$ as $N(y)$, one can find the normalized rank of item $\alpha$ by $R_\alpha(y)/N(y)$. To ensure a fair comparison, popularity is rescaled to keep the area under the popularity-rank curves the same for the results from different years. This can be achieved by computing the normalized item popularity ($D_\alpha$), given by $D_{\alpha}= \frac{k_\alpha(y)}{\sum_\beta{k_\beta(y)}}\div \frac{1}{N(y)}= \frac{k_\alpha(y)N(y)}{\sum_\beta{k_\beta(y)}}$, where $k_\alpha(y)$ is the degree of item $\alpha$ in year $y$.
\end{thebibliography}
\end{document}